\documentclass[useAMS,usenatbib]{mn2e}
\usepackage{graphicx}
\newcommand*\aap{A\&A}

\newcommand*\apj{ApJ}

\def\asec{\ifmmode ^{\prime\prime}\else$^{\prime\prime}$\fi}

\def\it{\sl}
\def\degs{\ifmmode ^{\circ}\else$^{\circ}$\fi}
\def\amin{\ifmmode ^{\prime}\else$^{\prime}$\fi}
\def\asec{\ifmmode ^{\prime\prime}\else$^{\prime\prime}$\fi}
\def\fm{\hbox{$.\!\!^{\rm m}$}}            
\def\farcs{\hbox{$.\!\!^{\prime\prime}$}}  

\def\psr{PSR~B1133+16}
\def\degs{\ifmmode ^{\circ}\else$^{\circ}$\fi}
\def\amin{\ifmmode ^{\prime}\else$^{\prime}$\fi}

\def\eqalign#1{\null\,\vcenter{\openup1\jot \m@th
   \ialign{\strut\hfil$\displaystyle{##}$&$\displaystyle{{}##}$\hfil
   \crcr#1\crcr}}\,}
   \title[On  the PSR\, B1133+16 optical counterpart]{On  the PSR\, B1133+16 optical counterpart.
   \thanks{Based on observations made with the Gran Telescopio Canarias (GTC), installed in the Spanish Observatorio del Roque de los Muchachos of the Instituto de Astrof'sica de Canarias, in the island of La Palma (program GTC1-12AMEX) and also on observations collected at the  European Organisation for Astronomical 
 Research in the Southern Hemisphere, Chile.  ESO number is 088.D-0298.}}

   \author[S. Zharikov and R. Mignani]{Sergey Zharikov$^{1}$,          Roberto P. Mignani$^{2,3,4}$ \\
$^{1}$  Observatorio Astronomico Nacional, Instituto de Astronomia, Universidad Nacional Autonoma de Mexico, Ensenada, BC, Mexico,\\
$^{2}$ INAF - Istituto di Astrofisica Spaziale e Fisica Cosmica Milano, via E. Bassini 15, 20133, Milano, Italy \\
$^{3}$ Mullard Space Science Laboratory, University College London, Holmbury St. Mary, Dorking, Surrey, RH5 6NT, UK \\
$^{4}$ Kepler Institute of Astronomy, University of Zielona G\'ora, Lubuska 2, 65-265, Zielona G\'ora, Poland}
 
\begin{document}
   \date{}
 \pagerange{\pageref{firstpage}--\pageref{lastpage}} \pubyear{2013}
\maketitle

\label{firstpage}
 \begin{abstract}
 The aim of this work is confirming the optical identification of \psr, whose candidate optical counterpart was detected in Very Large Telescope (VLT) images obtained back in 2003. We used new deep optical images of the \psr\ field obtained with both the 10.4 m Gran Telescopio Canarias  (GTC) and the VLT in the $g'$ and $B$ bands, respectively,  to confirm the detection of its candidate optical counterpart and its coincidence with the most recent pulsar's radio coordinates.  We did not detect any object at the position of the pulsar candidate counterpart ($B\sim28$), measured in our  2003 VLT images. However,  we  tentatively detected an object of comparable brightness in both the 2012 GTC and VLT images, whose position is offset by  $\sim 3\farcs03$ from that of the  pulsar's candidate counterpart  in the 2003 VLT images and lies along the pulsar's proper motion direction.  Accounting for the time span of $\sim9$ years between the 2012 quasi-contemporary  GTC and VLT images and the 2003 VLT one,  this offset is consistent  with  the  yearly displacement of the pulsar due to its proper motion.
  Therefore, both the flux of the object detected in the 2012 GTC and VLT images and its position, consistent with the proper motion-corrected pulsar radio coordinates, suggest that we have detected { the candidate pulsar   counterpart} that has moved away from its 2003 discovery position.  
\end{abstract}

\begin{keywords}
 pulsars:   general    --   pulsars,   individual:    PSR\, B1133+16  --
stars: neutron
\end{keywords}
 %

\section{Introduction}


PSR\,  B1133+16 is a $\sim$5Myr old, nearby  radio pulsar  at a parallactic distance of $\sim$350 pc, also detected in X-rays by  {\em Chandra}  \citep{2006ApJ...636..406K}. Recently, \cite{2008A&A...479..793Z} presented  the results of deep optical imaging of the {PSR\, B1133+16} field, taken  with  the ESO Very Large Telescope (VLT)  in  the $B$, $R_c$, and $H_\alpha$ bands in 2003 and in the $B$ band in 2001, aimed at searching for the optical counterpart of the pulsar and  its bow shock nebula.    A faint $B$=28\fm1(3) source  was detected and proposed as  the  optical counterpart  of PSR\, B1133+16,  since its position was consistent with that of the radio pulsar and its X-ray counterpart, measured by {\em Chandra}. The  upper limit in the $R$ band implied a colour index $B$-$R$$\la$0.5,  which is compatible with those of most pulsars  identified in the optical range \citep{2001A&A...376..213M, 2004A&A...417.1017Z, 2006A&A...448..313S}.  The pulsar radio parallax $\pi=2.80(16)$ mas measured  by \citet{2002ApJ...571..906B}  yields a distance of 350$\pm$20~pc. For this value, both the derived $B$-band optical luminosity  $L_B=5.75^{+2.76}_{-1.864} \times 10^{26}$ ergs cm$^{-2}$ s$^{-1}$  of the candidate counterpart and its ratio to the X-ray luminosity $L_X=5.01^{+4.30}_{-2.41}\times 10^{28}$  erg cm$^{-2}$ s$^{-1}$ in  the 2--10 keV energy range are also consistent  with  the expected values derived  from a sample of pulsars detected in both spectral domains \citep{2006AdSpR..37.1979Z}. Although the pulsar radio proper motion of $\mu_\alpha = -73.95(38)$ mas yr$^{-1}$ and  $\mu_\delta =  368.05(27)$ mas yr $^{-1}$ \citep{2002ApJ...571..906B}  yields a high transverse velocity of 631$\pm$30~km s$^{-1}$, no $H_\alpha$ Balmer bow shock  was detected by  \cite{2008A&A...479..793Z}, implying a low density  of  ambient matter around the pulsar.  However,  both in the  VLT $H_\alpha$ images and in the {\em Chandra} X-ray ones, they detected   the signature of a  trail  extending   $\sim 4\asec$---$5\asec$ behind the pulsar and coinciding with the direction of its proper motion.

The association of the proposed optical  counterpart with PSR\, B1133+16 can be easily verified by follow-up imaging of the field at a distance of a few years.  Owing to the pulsar proper motion, its candidate optical counterpart  is expected  to show an apparent displacement of $\sim 3\arcsec$ due north-west between our first-epoch VLT image (2003) and a second one acquired at the beginning of 2012.  The measurement of such a displacement would provide an unambiguous piece of evidence to confirm the proposed identification. In this paper, we report on the results of new observations of the pulsar field performed with the 8m VLT and the 10.4m Gran Telescopio Canarias (GTC) at the beginning of 2012, aimed at detecting the candidate optical counterpart at the proper motion-corrected pulsar position. 

The observations and  data reduction are described  in Sect.~\ref{sec2}, while the  results and conclusion are presented  and discussed in Sect.~\ref{sec3} and \ref{sec4}, respectively.

\begin{figure*}
\setlength{\unitlength}{1mm}
\resizebox{12cm}{!}{
\begin{picture}(120,150)(0,0)
\put (-32,0) {\includegraphics[width=9.cm,  bb = 20  0  900 850, clip]{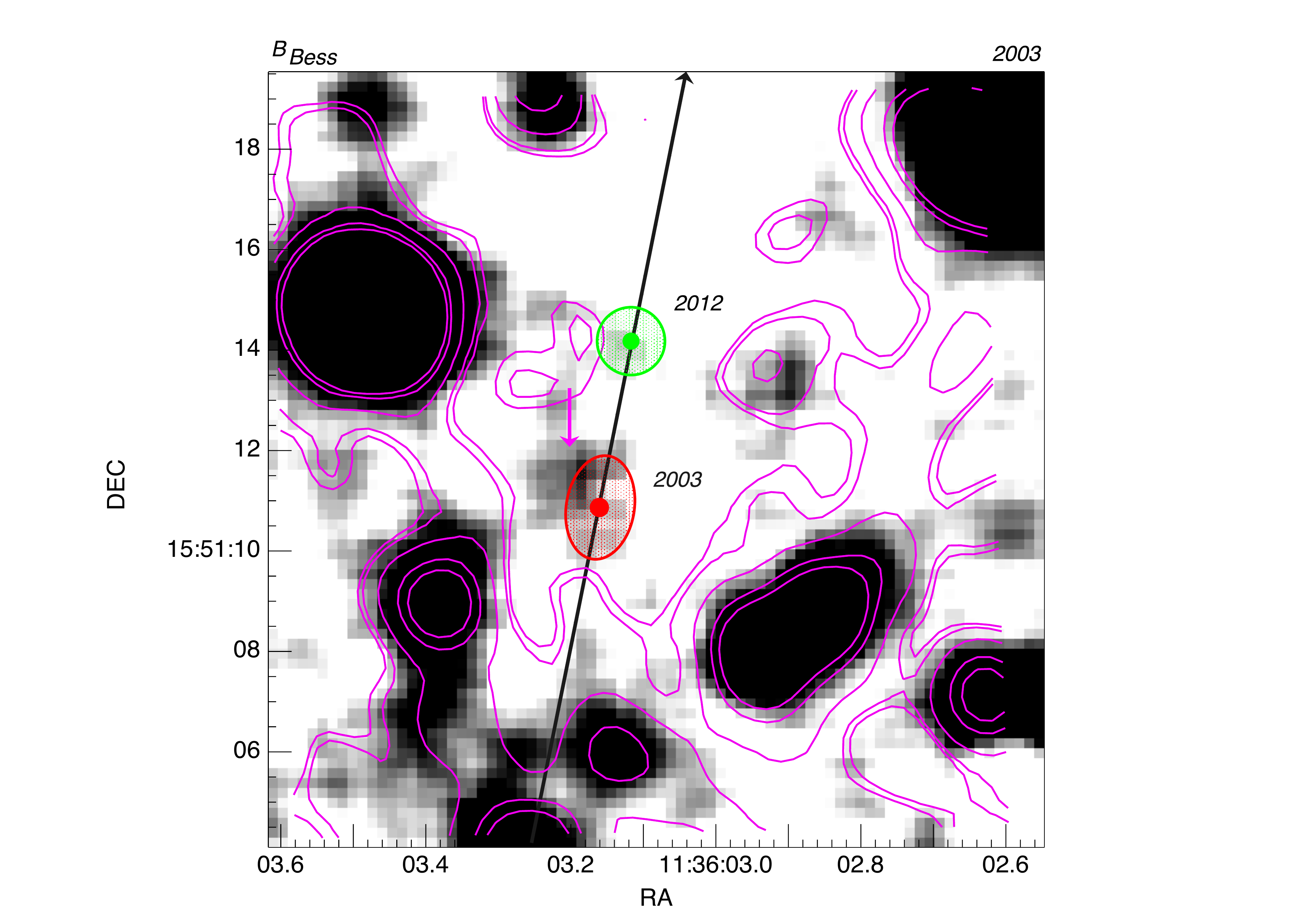}}
\put (-32,76) {\includegraphics[width=10.35cm, bb = 40  -10  900 850, clip]{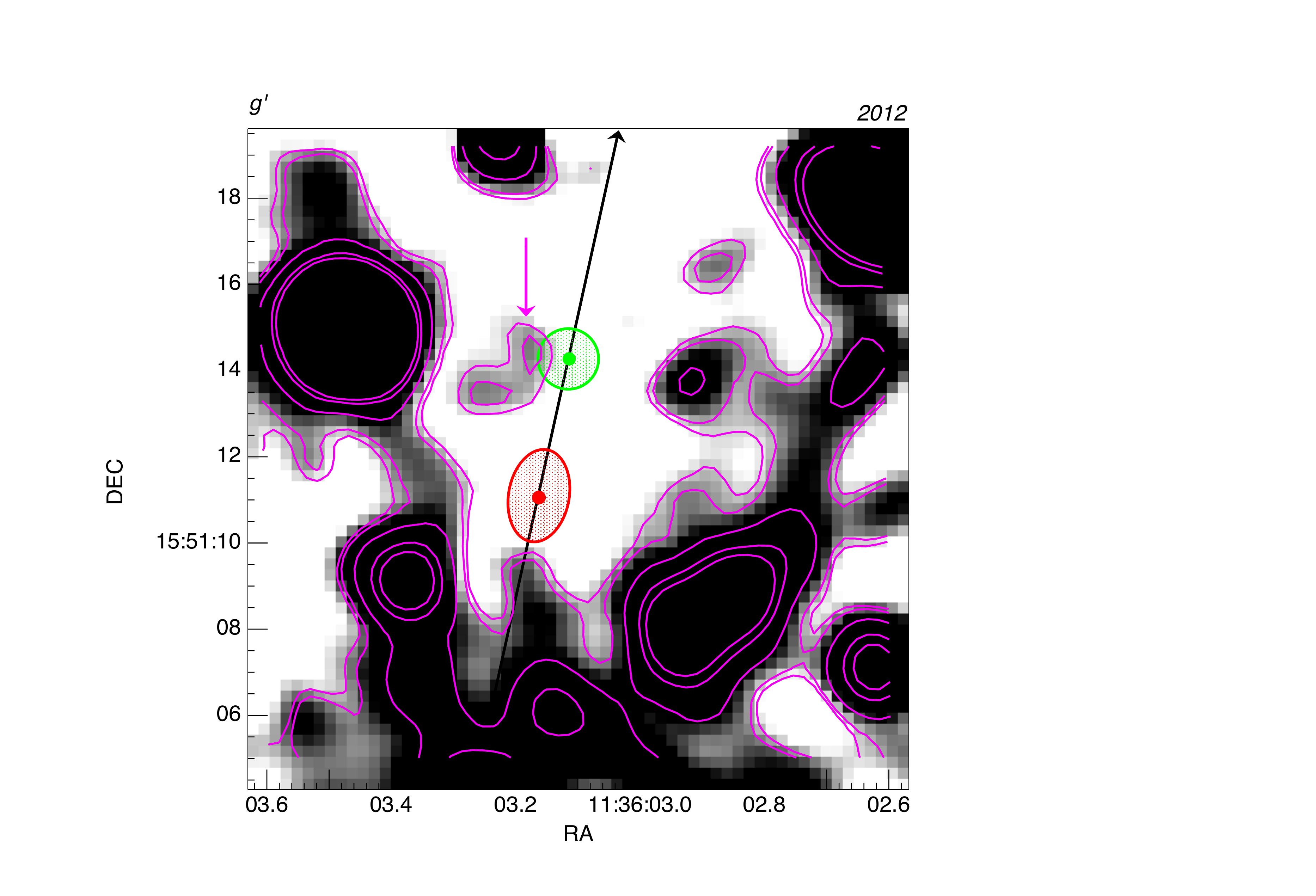}}
\put (53,73) {\includegraphics[width=10cm, bb = 60 -20 900 810,clip]{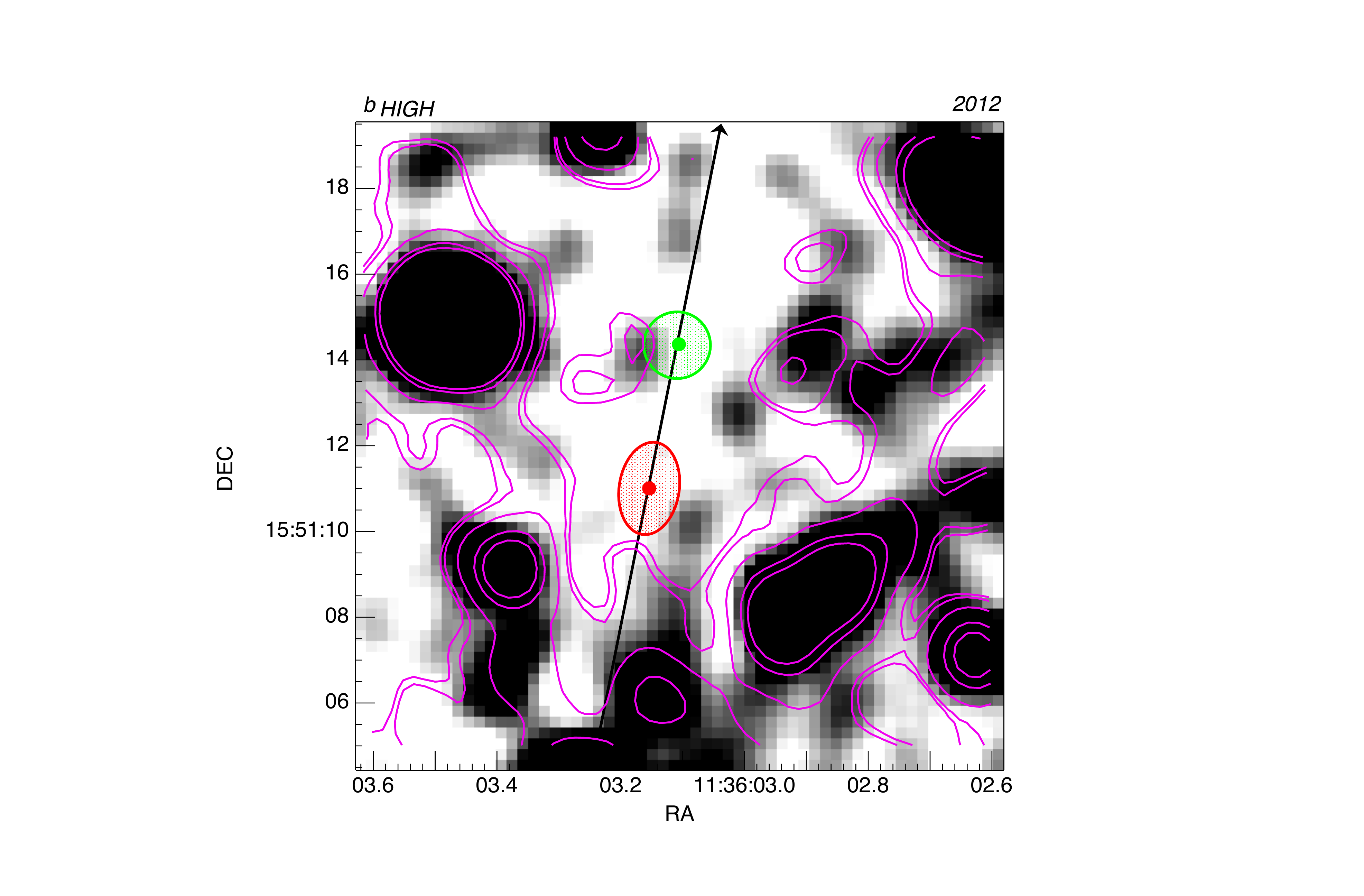}}
\put (55,-5) {\includegraphics[width=10.15cm,bb = 60 0 900 810,clip]{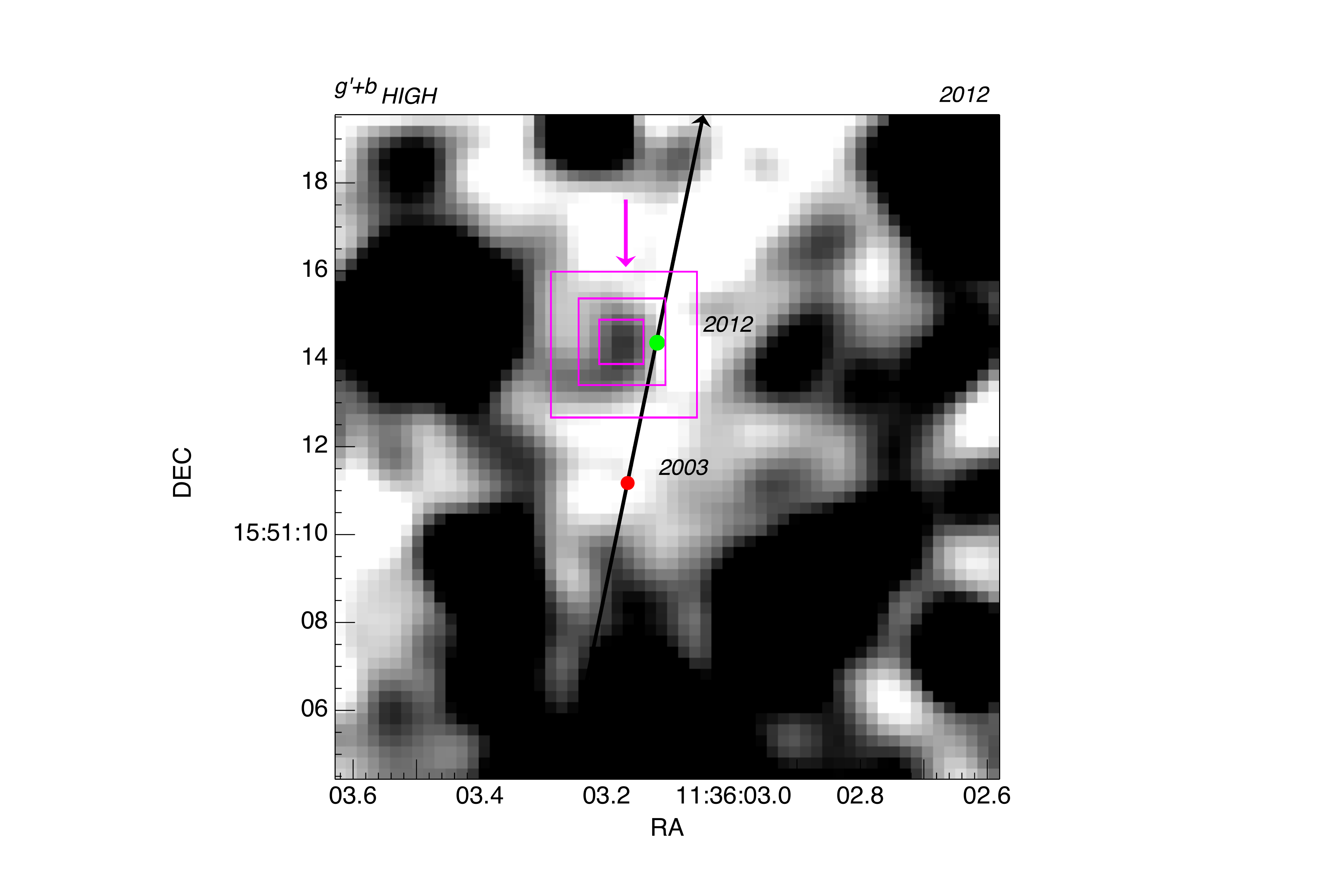}}
\end{picture}}
  \caption{{\em Top:} 15\arcsec $\times$ 15\arcsec\ sections of the 2012 GTC/OSIRIS $g'$ (10967 s; left panel) and VLT/FORS2 $b_{\rm HIGH}$-band  (9000 s; right panel) images of  the PSR\, B1133+16 field.  For a better representation, both images have been smoothed with a $3\times3$ pixels Gaussian kernel that approximately   corresponds to the size of the measured PSF.  
  {\em Bottom:} 15\arcsec $\times$ 15\arcsec\ section of the 2003 VLT/FORS2 $B_{\rm BESS}$-band image of 
  Zharikov et al.\ 2008a
  (left panel) and co-addition of the 2012 GTC/OSIRIS and VLT/FORS2 $g'$ and $b_{\rm HIGH}$-band images (right panel). 
  The magenta contours from the GTC $g'$-band image are overlaid on the 2012 VLT/FORS2 $b_{\rm HIGH}$-band  and 2003 $B_{BESS}$ images, as a reference. { The first contour corresponds to  the flux from an object detected  in the  $ 1\arcsec \times 1\arcsec$ aperture with  the $\sim 3\sigma$ level. Others contour have been  selected arbitrary for the best presentation.}
    We note that the pixel size of the 2003 VLT image was twice as small as that of the 2012 VLT one. In all images,  coloured dots  mark the position of the pulsar at the epoch of the 2003 VLT observation (red) and  the 2012 GTC and VLT observations (green).  The radio interferometric positions are known to negligible uncertainties on this figure.    The object marked by the arrow at the 2003 pulsar position in the lower left panel is { the  candidate pulsar counterpart} of Zharikov et al.\ 2008a, whereas the arrow in the other panels marks the proposed counterpart at the 2012 pulsar position.  The ellipse and the circle mark the $3\sigma$ error boxes on { the pulsar's radio positions}, which follow from the accuracy of the astrometry calibration of the optical images, the error on the pulsar radio coordinates, and the error on the pulsar's radio proper motion extrapolation, while the black arrow  shows the direction of the pulsar proper motion. The size and the shape  of the red ellipse takes in account the time span of the observations during the 2003 VLT runs (Zharikov et al.\ 2008a). The magenta rectangles in the bottom-right panel show the integrating aperture $1^{\prime\prime}\times1^{\prime\prime}$, the ignored area, and the local background area used for photometric measurements.
     } 
 
 \label{fig1}
\end{figure*}

\section{Observation and data reduction}
\label{sec2}

\begin{table}
\caption{Log of the VLT/FORS2 and GTC/OSIRIS observations of \psr\ (first and second half, respectively). }
\begin{tabular}{lcllcl}
\hline\hline
 Date              &  Filter  &  $T_{exp}$      &    AM$^*$      & Seeing$^*$       \\
    (yyyy-mm-dd)             &     & Exposure(s)  &           Airmass        & ($\arcsec$)     \\  \hline \hline                 
    2012-03-25     &  $b_{\rm HIGH}$&$600\times5$      &  1.35(5)        &    0.70(5)         \\
    2012-03-26     &  $b_{\rm HIGH}$&$600\times10$      &  1.35(5)         &    0.72(5)           \\ \hline
    2012-04-22      &  $g'$&$ 686\times16$      &  1.13(10)       &    1.05(10)                       \\                               
   \hline
\end{tabular}
\begin{tabular}{l}
 $^*$ - {average value during the run and its  dispersion.}
\end{tabular}
\label{t:log}
\end{table}

\begin{figure*}
\setlength{\unitlength}{1mm}
\resizebox{12.cm}{!}{
\begin{picture}(120,70)(0,0)
\put (-30,0) {\includegraphics[width=8.2cm, bb = 0  0 550 550, clip]{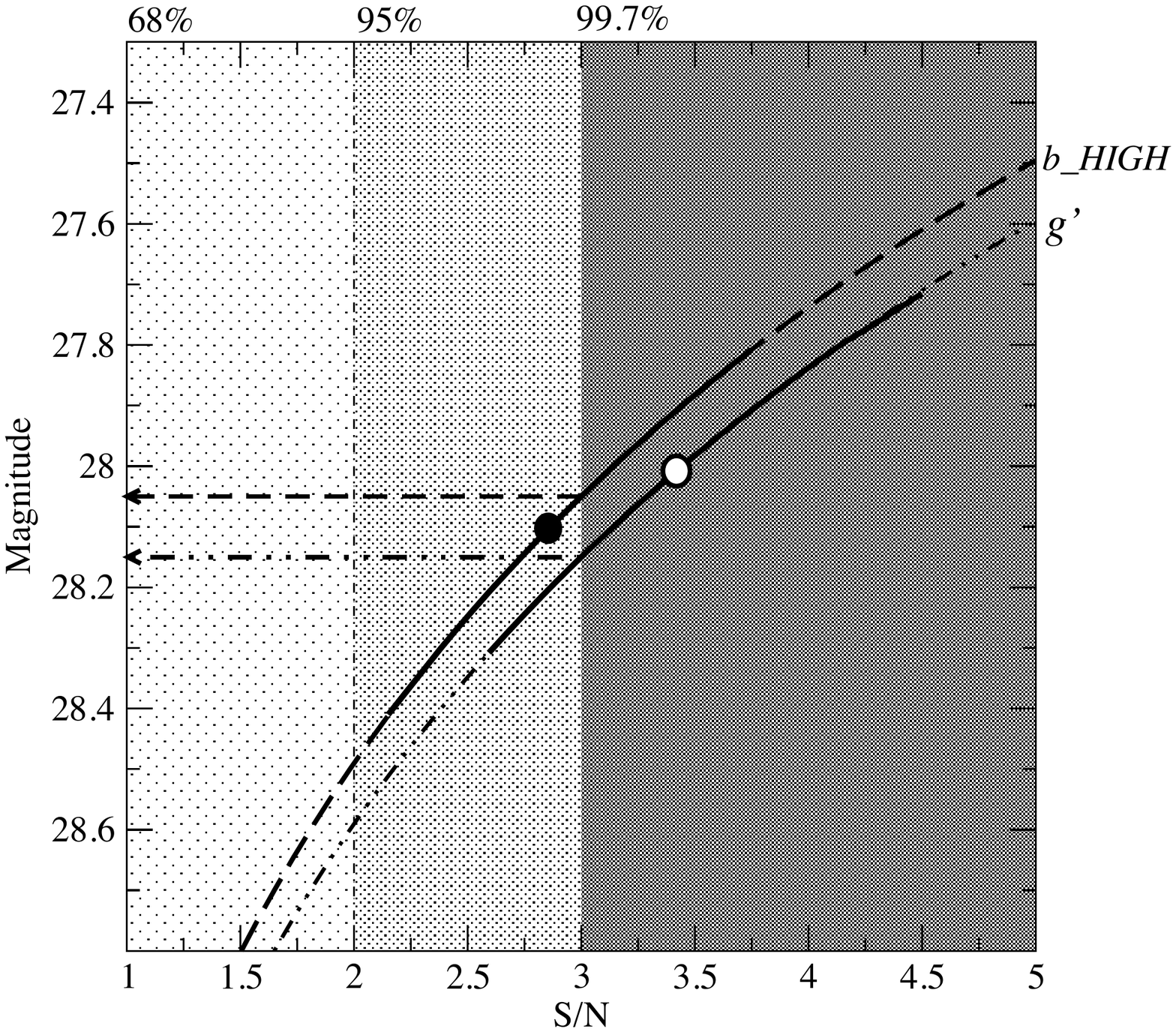}}
\put (65,0) {\includegraphics[width=8.2cm, bb = 0 0 550 550, clip]{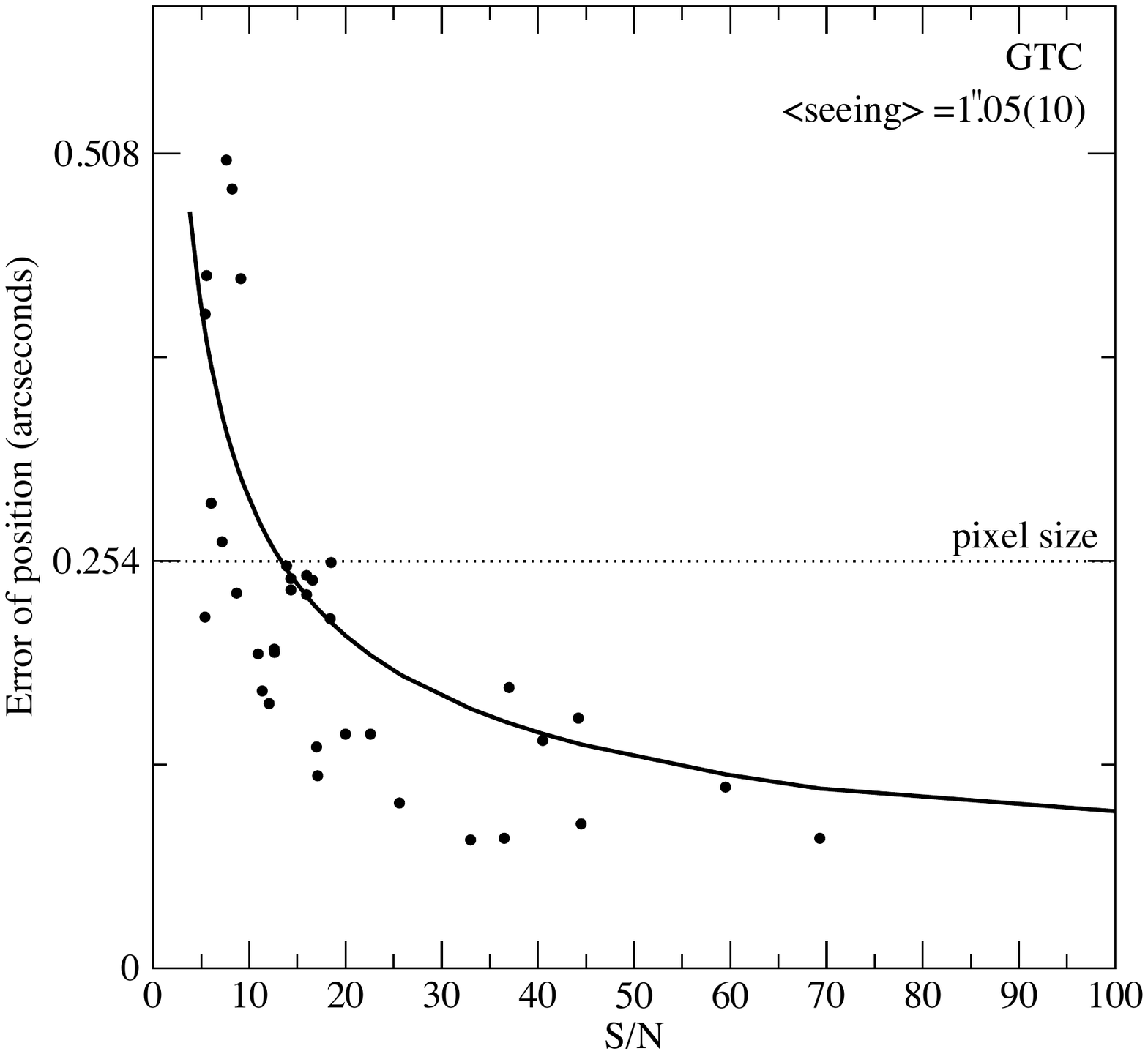}}
\end{picture}}
  \caption{{\em Left panel:} Magnitude vs. signal--to--noise (SN) ratio calculated for both the co-added 2012 VLT $b_{\rm HIGH}$  and GTC  $g'$ images. Magnitudes and associated errors  of the \psr\ candidate counterpart are shown by the circles and thick solid lines, respectively. The numbers on the top axis show the detection significance corresponding to a given S/N ratio.  The arrows mark the computed $3\sigma$ detection level in each band. {\em Right panel:}  Dependence of position errors vs. S/N ratio for  a sample of objects  selected at random in the GTC/OSIRIS field of view. The solid line corresponds to the fit to the measured points with a function  $y=A \times x^B$, which gives A=0.92 and B = -0.5. The error on the displacement of an object with S/N ratio as low as $\sim 3$, such as the \psr\ candidate counterpart, can be as high as half of the measured image seeing.  }
 \label{fig2}
\end{figure*}

We observed the field of \psr\  using  the FOcal Reducer/low dispersion Spectrograph,  (FORS2; \cite{1998Msngr..94....1A})
at the VLT (program 088.D-0298(A)), and the Optical System for Imaging and low Resolution Integrated Spectroscopy (OSIRIS; \cite{2008SPIE.7014E.107R}) at the GTC  (program GTC1-12AMEX) through the  $b_{\rm HIGH}$  ($\lambda$ = 4400$\pm517$ \AA)  and $g'$ ($\lambda$ = 4815$\pm765$ \AA) filters, respectively. The total integration time was 9000s in the $b_{\rm HIGH}$ and 10967s in the  $g'$ filters. To increase the signal--to--noise, for both the VLT and GTC observations we used a  $2\times2$ binning of the detector, which gives a  projected pixel size on the sky of 0\farcs252 and 0\farcs254, respectively. The observation log is summarised in Table \ref{t:log}. Images of standard stars and day/twilight calibration frames were taken in each observing run. We reduced the science images in a standard way  using dedicated programs in the {\sc iraf} and {\sc midas} packages  to remove instrumental effects, i.e.  by applying the bias and flat-field correction, and removed the cosmic rays' tracks before image alignment and stacking.   We removed the residual heterogeneity of the sky  background  in the vicinity ($200\times300$ pixels or 50\farcs8$\times$75\farcs2) of the pulsar in GTC images using the {\sc midas} command {\tt fit/flat\_sky} with a second order two-dimensional polynomial.

 We used the astrometric solution obtained by \cite{2008A&A...479..793Z} for the 2003 VLT/FORS2 images to apply the 
 astrometric calibration to the 2012 GTC and VLT ones.  To do this, we measured the pixel and absolute coordinates of five stars in vicinity of the expected pulsar position and applied the {\tt ccmap} task in  {\sc iraf} to compute the pixel--to--sky coordinate transformation in the new GTC and VLT images. The formal rms of the fit is very small ($\sim$0\farcs01) because of the small region ($200\times300$ pixels) and number of reference stars that we used. Enlarging the region and including more reference stars did not change this result significantly. Therefore, the errors of the astrometric solution of the 2003 VLT/FORS2 image ($1\sigma \sim$ 0\farcs2  in RA and DEC) 
 dominate  the overall astrometric accuracy of the new data.
  
  We performed the  photometric calibration using standard stars in the NGC\, 2298 and Leo\, I fields  (VLT/FORS2) and in the PG\, 0918+029D one (GTC/OSIRIS) observed during the nights. We applied the airmass correction using the most recent extinction coefficients $k_i$  for Cerro Paranal and Roque de Los Muchachos, available on the observatories web pages\footnote{
  http://www.eso.org/observing/dfo/quality/FORS2/qc/ \\ photcoeff/photcoeffs\_fors2.html \\
  http://www.ing.iac.es/Astronomy/observing/manuals/ \\ ps/tech\_notes/tn031.pdf 
  }.
The computed zero-point of  the 2012 VLT/FORS2 image is $C(b_{\rm HIGH}$) = 27.83(3) (in ADU).   We determined the formal $3\sigma$ detection limit (99.7\% significance level) of the image as the flux of a star-like object corresponding to $3\sigma_{bg}$, where $\sigma_{bg} = \sqrt{F^{bg}}$  and $F^{bg}$ is
a flux of  the sky background
integrated in  a rectangular  aperture of 1\farcs0$\times$1\farcs0 size.
We chose the size and shape\footnote{ The rectangular shape of the aperture was selected by us to avoid additional uncertainties following from approximating the area of a circular aperture filled by rectangular pixels.} of aperture to maximise the S/N ratio, according to the seeing condition during the observations (Table \ref{t:log}).  For consistency with the standard stars' photometry, the measured flux was corrected for the finite size of the aperture  using the computed aperture correction $\delta m_i$. By applying the computed zero point and the atmospheric extinction correction,  we derived $3\sigma$ detection limits:
$$ m_i^{3\sigma} = C_i -2.5\log\left(\frac{3\sqrt{F^{bg}_i}}{T_{exp}}\right) -\delta m_i- k_i \times AM; i  = g', b_{HIGH} $$
that correspond to  $b_{\rm HIGH} \approx 28.05 = 2.4\times10^{-2}\mu Jy$.
 We computed the Sloan $g'$ magnitude of the Landolt's standard star PG\, 0918+029D using the relation  $g' = V +0.54(B-V)-0.07$ from \cite[see Table~7]{2002AJ....123.2121S}, which yields a zero-point $C(g')=28.74(2)$ for the GTC/OSIRIS image, the same as listed in the instrument web page. By applying this value, and using the same approach as above, we estimated the corresponding  formal $3\sigma$ detection limit as $g'\approx28.15 = 2.0\times10^{-2}\mu Jy$.
 The detection limits are comparable in deepness for both telescopes/instruments, with the larger aperture of the GTC compensated by the better seeing conditions of the VLT observations (see Table \ref{t:log}).

\begin{figure*}
\setlength{\unitlength}{1mm}
\resizebox{12.cm}{!}{
\begin{picture}(120,150)(0,0)
\put (-42,70) {\includegraphics[width=10.45 cm, bb = 0 0 900 900,clip]{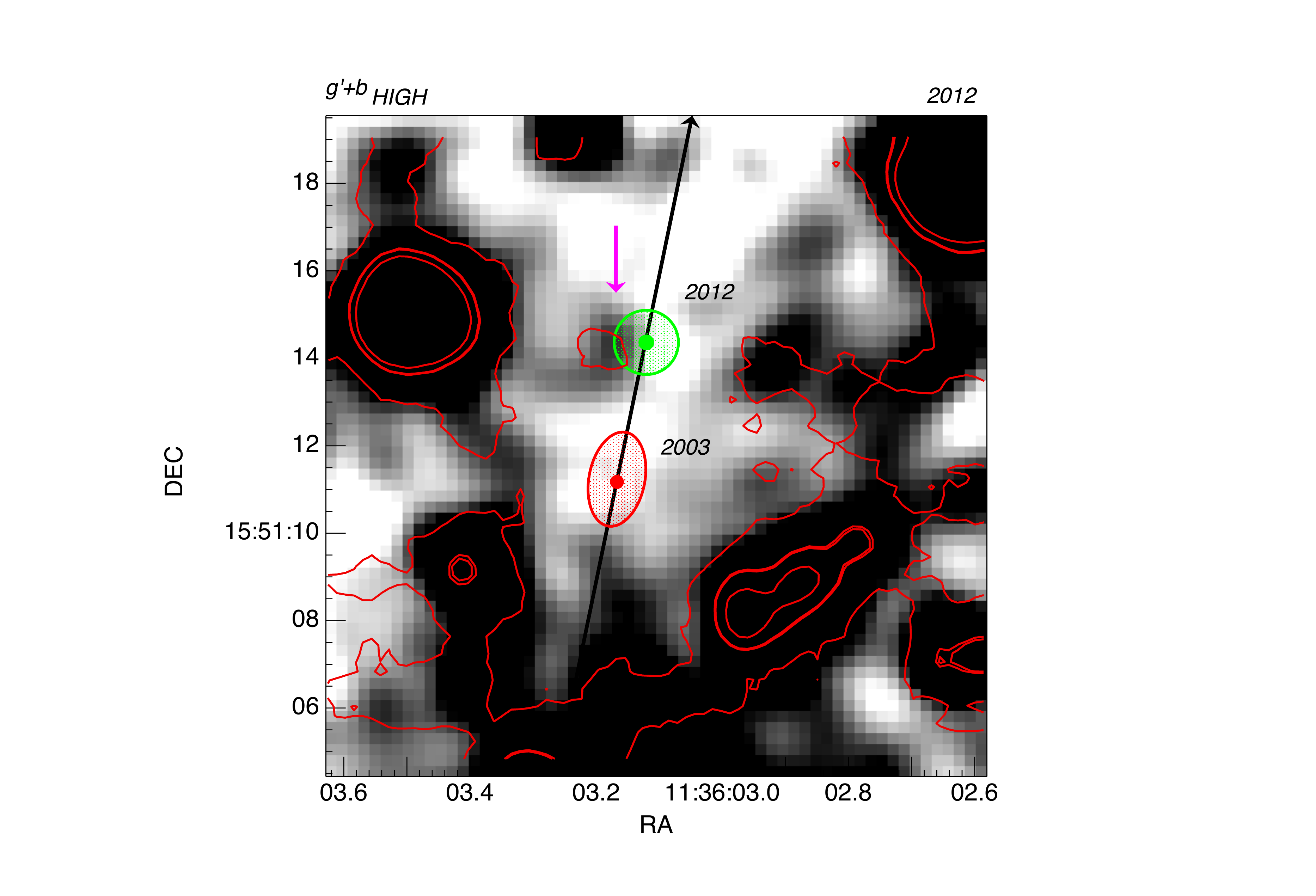}}
\put (-33,3) {\includegraphics[width=9 cm, bb =  0 0 900 900,clip]{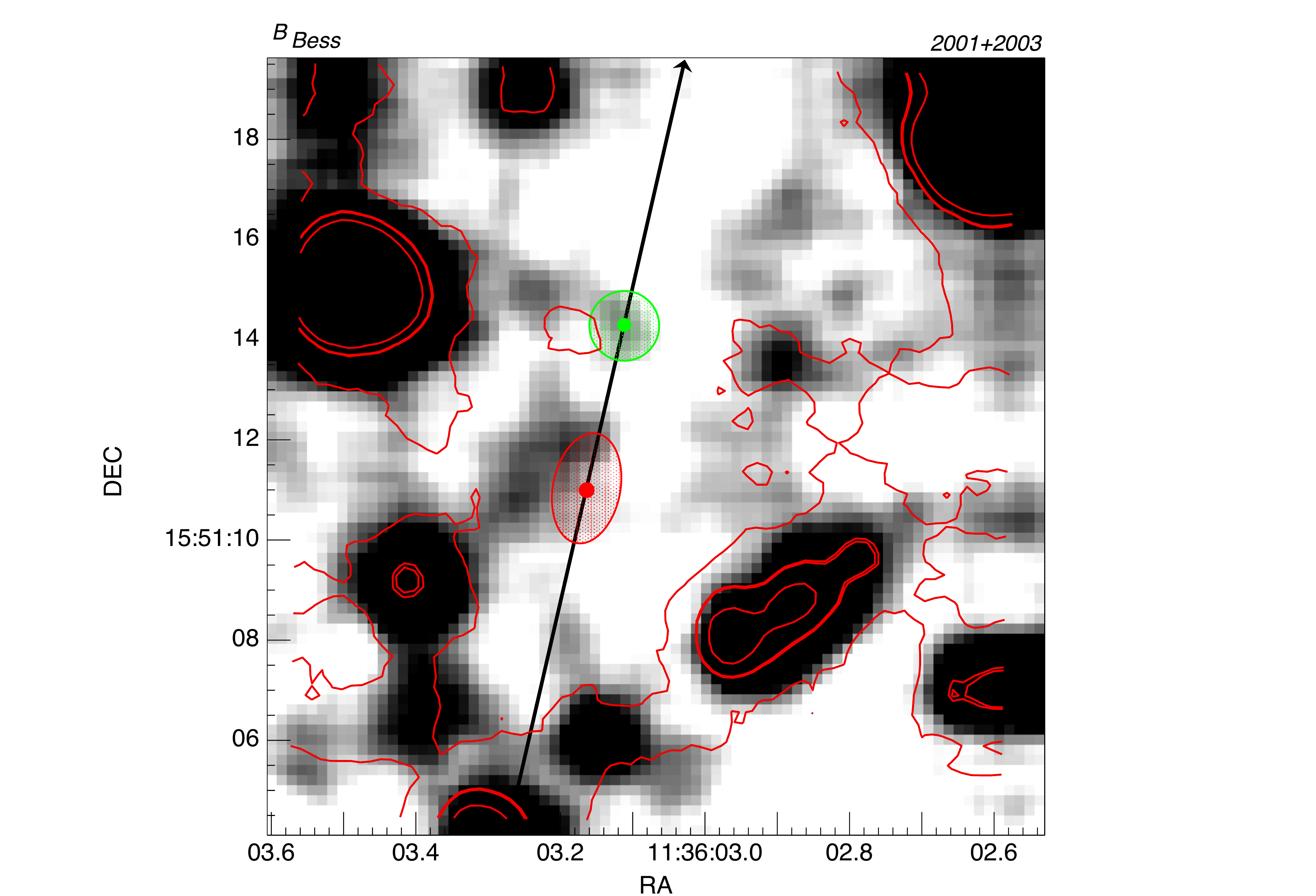}}
\put (55,3) {\includegraphics[width=9 cm, bb = 0 0 900 900,clip]{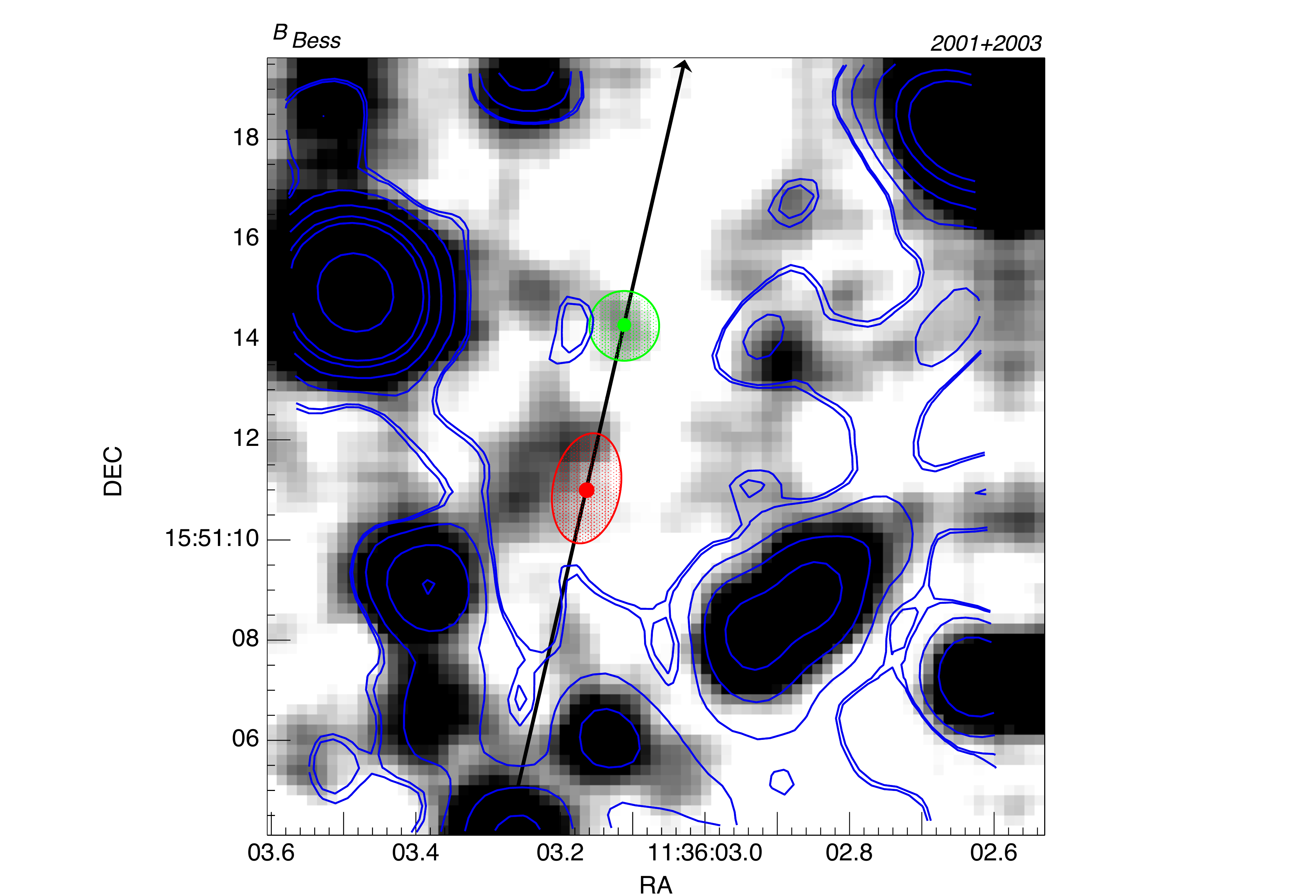}}
\put (54,76) {\includegraphics[width=8.8 cm, bb = 0 0 900 900,clip]{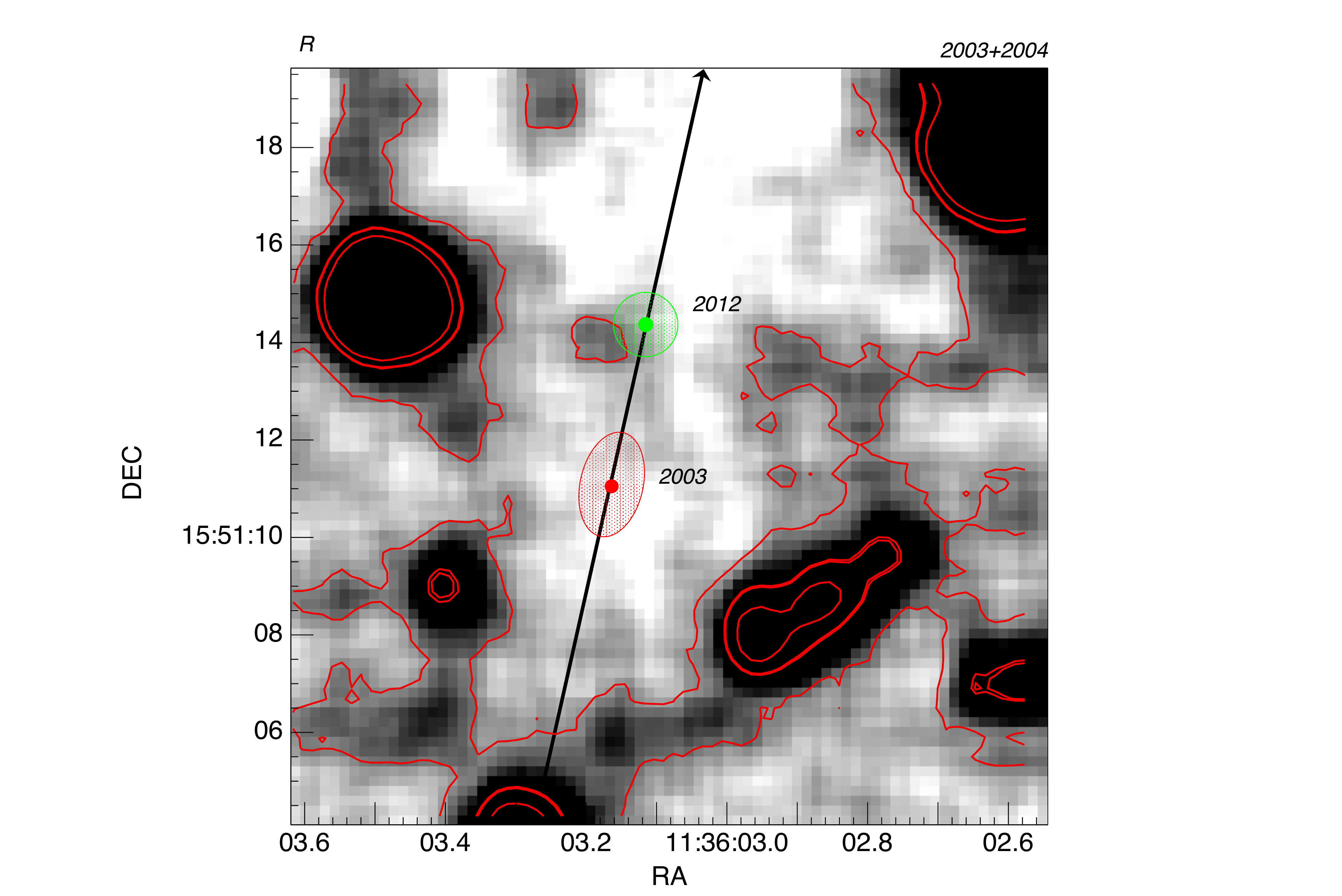}}
\end{picture}}
  \caption{15\arcsec $\times$ 15\arcsec\ sections of the \psr\ field. Images have been smoothed by  a $3\times3$ pixel Gaussian kernel. The panels show the co-addition of the 2012 VLT/FORS2 and GTC/OSIRIS  images, obtained through the  $b_{\rm HIGH}$ and $g'$ filters, respectively (top left), the co-addition of the archival VLT/FORS2 $R$-band images taken in 2003 and 2004 (top right), the co-addition of the $B$-band images of Zharikov et al. 2008  taken in 2001 and 2003 (bottom panels).   The red contours correspond to the 2003+2004 VLT $R$-band image, while the blue contours (bottom right panel) correspond to the co-addition of the 2012 VLT and GTC  images.   As in Fig. 1, the red and green dots mark the 2003 and 2012 { pulsar's radio positions}, respectively,  with their associated $3\sigma$ uncertainties marked by the red ellipse and the green circle, respectively. The black arrow shows the direction of the pulsar's proper motion. { Contours have been selected also in the same way as in Fig.\ref{fig1}.}  
  }
 \label{fig3}
\end{figure*}

\section{Results}
\label{sec3}

The results of our observations are presented in Fig. \ref{fig1}. The top two panels show the co-added 2012 GTC/OSIRIS $g'$(left) and VLT/FORS2 $b_{\rm HIGH}$-band (right) images, respectively. The stack of the two images is shown in the bottom right 
panel. The bottom left panel shows, as a reference, the 2003 VLT/FORS2 image reported by \cite{2008A&A...479..793Z} and taken through the $B_{\rm BESS}$ filter  ($\lambda$ = 4290$\pm440$ \AA). 
 In all images,  colored dots mark  the expected position of the pulsar at the epoch of the 2003 VLT observation (red) and 2012 VLT$+$GTC observations (green), computed according to its measured proper motion \citep{2002ApJ...571..906B}. 
The 2003 and 2012 pulsar coordinates are summarised in  Table \ref{t:cor}.
 The  red ellipse and green  circle  mark the $3\sigma$ error boxes associated with the 2003 and 2012 pulsar positions (Table \ref{t:cor}), which follow from the accuracy of the astrometry calibration of the optical images, the error on the pulsar radio coordinates  at the reference epoch, and the error on the pulsar's radio proper motion extrapolation.   The fact that the uncertainty on the pulsar position is larger at the epoch of the 2003 observations is due to the fact that they were spread over a larger time span than the 2012 ones, which increases the uncertainty due to the propagation of the proper motion error. The black  arrow  shows the direction of the pulsar proper motion.  The candidate optical counterpart to \psr, detected in both the 2003 and 2012 images, is tagged by the vertical arrows.

\begin{table}
\caption{Expected coordinates of \psr\ at the epochs of the 2003 and 2012 VLT observations (second and third row) computed according to its proper motion \citep{2002ApJ...571..906B}.  The pulsar coordinates given by \citet{2002ApJ...571..906B} are reported in the first row. The associated $3\sigma$ error in both right ascension and declination is 0\farcs046. Columns $\delta$RA and $\delta$DEC corresponds to the $3 \sigma$ uncertainties on the expected position in right ascension and declination, which follow both from f the accuracy on the reference radio coordinates, the propagation of the proper motion error, and the accuracy of the astrometry calibration of the optical images. The fact that the uncertainty is larger at the epoch of the 2003 observations is due to the fact that they were spread over a larger time span than the 2012 ones.   }
\begin{tabular}{lllll} \hline \hline
Epoch & RA & DEC & $\delta$RA & $\delta$DEC \\
   & hh mm ss & $^o$~ '~ " & (\arcsec) & (\arcsec)  \\ \hline \hline
 2000.00 & 11 36 03.183 & +15 51 09.726 &  0.046$^\dagger$  & 0.046$^\dagger$  \\
 2003.66  & 11 36 03.16 & +15 51 11.06 &  0.63$^{\star}$ & 1.02$^{\star}$ \\   
 2012.23 &  11 36 03.11 & +15 51 14.37 & 0.64 & 0.64\\   
\hline
\end{tabular}
\begin{tabular}{l}
$^\dagger$ \citet{2002ApJ...571..906B} \\
$^\star$  Zharikov et al.\ (2008a) \\ 
\end{tabular}
\label{t:cor}
\end{table}

As seen, the pulsar's candidate counterpart of  \cite{2008A&A...479..793Z} is not detected at its 2003 discovery position in our new images. However, an object is marginally detected near the expected  2012 pulsar radio position in both the GTC $g'$ and VLT $b_{\rm HIGH}$-band images (Fig.\ref{fig1}, top left and right panel, respectively). The shape of the object is consistent with a point-like source in the $b_{\rm HIGH}$-band image but it looks somehow extended in the $g'$-band one, possibly owing to a blending with a nearby object or background fluctuation and the worse seeing conditions of the GTC observations.

  We measured its magnitude through aperture photometry using the same approach as in \cite{2008A&A...479..793Z},  and as described in Sectn. 2, and choosing an optimal aperture of  $4\times4$ pixels ($\sim 1\arcsec \times 1\arcsec$)  (see Fig. ~\ref{fig1}, lower right panel).  The object's fluxes are $g'=28.0(3)$ and $b_{\rm HIGH}=28.1(3)$ (see footnote to Table~\ref{t:mag} for details of the measurement), where the associated errors are purely statistical.    We note that,  after accounting for the differences between the different filters, the measured object's fluxes 
  are comparable with that of the {candidate pulsar  counterpart} ($B_{\rm BESS}=28.1(3)$; Fig.\ref{fig1}, lower left panel) discovered by \cite{2008A&A...479..793Z}. Although in both  the 2012 GTC and VLT images the object's signal--to--noise (S/N) ratio corresponds to  a $\sim 3 \sigma$ detection only (Fig.\ref{fig2}, left),  its independent detection  in both  of them suggests that it is real and not a background fluctuation. The object detection is also made more apparent by the co-addition of the GTC and VLT images (Fig.\ref{fig1}, lower right panel).  However, to rule out the possibility that this object is spurious, we  estimated the number of counts produced by random fluctuations of the background at random positions selected within a radius of   $3^{\prime\prime}$ around the object's position and using the same aperture as used to measure its flux. The brightest background fluctuations in a box of $4\times4$ pixels ($1\arcsec \times 1\arcsec$) only give  fluxes of $g' \sim 29.1$ and $b_{\rm HIGH} \sim 28.8$. This means that the flux of the object detected in both the GTC and VLT images cannot be due to background fluctuations but  it is likely ascribed to a real astrophysical  source.

  
 The position of the object detected in the 2012 GTC and VLT images (Fig.\ref{fig1}, top left and right panel, respectively) is slightly offset to the East ($\Delta r \sim $0\farcs6) from the expected 2012 pulsar position (green  dot), although such an offset is still consistent with the overall $3\sigma$ error on the computed pulsar's position (green hatched  circle). The same is true for the pulsar candidate counterpart detected in the 2003 VLT image (Fig.\ref{fig1}, bottom left), whose position is slightly offset to North East with respect to the computed pulsar position at the same epoch (red  dot) but still within the associated uncertainty (red hatched  ellipse). 
 Interestingly enough, the offset between the object detected in the 2012 VLT and GTC images and the pulsar candidate counterpart   detected in the 2003 VLT image (3\farcs03) is in the direction of the pulsar proper motion.  Such an offset differs by only  0\farcs27 from the computed pulsar displacement  (3\farcs3) between the epochs of the 2003 VLT and the 2012 VLT and GTC observations. 
 To estimate the significance of this difference we used the GTC images (686 s) to select a sample of objects detected at various S/N in the pulsar vicinity.  For each image, we measured the objects' positions on the detector and calculated the associated uncertainties as a function of the S/N ratio (Fig.\ref{fig2}, right).  As expected, the position uncertainties increase with decreasing of the S/N ratio and are as large 2 pixels (0\farcs56), i.e about half of the image PSF,  for objects detected with $S/N < 10$. In particular,  the 0\farcs27 difference between the 2003--2012 pulsar displacement and the offset of the object detected in the 2012 images from the pulsar candidate counterpart in the 2003 one does not exceed the estimated  position uncertainty  for an object detected with $S/N \sim 3$. 
 
Therefore, within the position errors,  we can claim coincidence between the computed 2012 radio position of \psr\ and that of the object detected in the 2012 GTC and VLT images.  Moreover, we can also claim consistency between the offset of the object position from that of the pulsar candidate counterpart of \cite{2008A&A...479..793Z} detected in the 2003 VLT image and the pulsar's proper motion displacement between the two epochs. This, together with the similar brightness of the two objects, suggests that  we detected the \psr\  candidate counterpart of \cite{2008A&A...479..793Z} in the 2012 GTC and VLT images, which is not detected at its 2003 discovery position (Fig.\ref{fig1}, top right and left panel) as the result of the pulsar proper motion.

Of course,  another possibility is that the  candidate counterpart of  \cite{2008A&A...479..793Z} has not moved from its 2003 discovery position, hence is not the actual pulsar counterpart, and  has not been detected in our more recent data just because its flux ($B_{\rm BESS}=28.1(3)$) is comparable with the $3\sigma$  detection limits of both the 2012 GTC ($g'\approx28.15$) and VLT ($b_{\rm HIGH} \approx 28.05$) images (Fig.\ref{fig2}, left).  In this case, the  object  found at the 2012 pulsar's radio position should have been visible in our previous images of the field.  
We show in Fig. \ref{fig3}  the co-added 2012 GTC $g'$ and VLT $b_{\rm HIGH}$-band image of the \psr\ field  (top left panel) and the co-added 2003+2004 VLT $R$-band one (top right panel) presented in  \cite{2008A&A...479..793Z}.  In both panels, the contours of the $R$-band image are overlaid (in red).  As seen, a faint object, of magnitude $R=26.8(2)$, is detected in the $R$-band image, very close to the computed 2012 pulsar's  radio position and that of the object detected in the co-added 2012 GTC $g'$ and VLT $b_{\rm HIGH}$-band images.  Thus, it is possible that the two are the same object, which means that  the latter would not  be the pulsar's counterpart.   To investigate this possibility we checked whether the two objects are also detected in the co-added 2001+2003 VLT $B_{\rm BESS}$-band image. As seen, although there are a few obvious background fluctuations within 1\arcsec\ from the positions of the two objects, no object is clearly detected  in this region (Fig. \ref{fig3}, lower left and right). We note that the $3\sigma$ detection limit of the 2001+2003 VLT $B_{\rm BESS}$-band image is $B_{\rm BESS}=28.6$ \citep{2008A&A...479..793Z}, which would correspond to an unusually red colour $B-R \ga 1.8$ for the object detected in the 2003+2004 VLT $R$-band image.  Moreover, the $B_{\rm BESS}$-band limit  is significantly fainter than both the 2012 GTC and VLT $3\sigma$ detection limits ($g'\approx28.15$; $b_{\rm HIGH} \approx 28.05$). This means that the object detected in the 2003+2004 VLT $R$-band image, but undetected in 2001+2003  VLT $B_{\rm BESS}$-band one, cannot be the same as that detected in the co-added 2012  GTC $g'$ and VLT $b_{\rm HIGH}$-band images. 
 
  Another possibility is that the object detected in  the 2003+2004 VLT R-band image
is variable. In this case, it might have been undetected in the 2001+2003 VLT $B_{\rm BESS}$-band image but detected in the 2012 GTC $g'$ and VLT $b_{\rm HIGH}$-band one.
 Unfortunately, the lack of an  R-band observation in 2012 for comparison with the 2003+2004 VLT one does not allow us to directly rule out this possibility. However, we estimate that this probability is very low since we did not find any faint but strongly variable object in the pulsar vicinity.

 Finally, one possibility is that the objects detected at the pulsar radio positions in the 2003 and 2012 images are both background fluctuations. \cite{2008A&A...479..793Z} estimated that the probability of detecting a source  at the $3\sigma$ level at the 2003 radio pulsar position was about 60-90\%, as follows from the number density of faint sources in the pulsar vicinity. This means that the probability of not detecting the same $3\sigma$ source in an independent set of observations, taken in the same pass band, with approximately equal conditions and exposure lengths, as the result of background fluctuations is about 10\%--40\%. 
Accounting for the difference in pass bands and weather conditions, we conservatively estimate a  maximum probability of 50\%.
Therefore, the probability of not detecting the \psr\ candidate counterpart, detected in the 2003 VLT B-band image, in both the 2012  VLT and GTC ones is $<$25\%. The probability of detecting a spurious source at the 2012 pulsar position in both the VLT and GTC images is $<$25\%, too. The combined probability is thus $\la$ 6\% and, taking into account the absence of any source at the 2012 pulsar position in the VLT 2001+2003 R-band images, this probability decreases to $\la$3\%.




  To summarise,  we think that is plausible that the object that we tentatively detected at the expected pulsar position  in the 2012 GTC and VLT images is  indeed { the candidate pulsar  counterpart} of \cite{2008A&A...479..793Z} that has moved along the pulsar's proper motion direction.
 

\begin{table}
\begin{center}
\caption{Measured $g'$ and $b\_{\rm HIGH}$ magnitudes and fluxes of  the \psr\  candidate counterpart,  computed through aperture photometry.}
\begin{tabular}{lccccll}
\hline\hline
 Telescope   & Flux$^{ap}$     & $k$     & $\delta m$   & $m$ & log(F)    \\
     filter                                    &  (ADU/s)               &     &                   &                &   $\mu Jy$  \\\hline \hline                 
    GTC/$g'$      &  0.752         & 0.14     &    0.83         &  28.0(3)     &   -1.62(12)             \\       
    VLT/$b_{\rm HIGH}$ &  0.356          & 0.25     &    0.57        &  28.1(3)    &    -1.64(12)             \\ \hline
 \end{tabular}\\
 \begin{tabular}{ll}
  Flux$^{ap} =$ (Flux$_{1^{\prime\prime}\times1^{\prime\prime}} - $Flux$^{SKY}_{1^{\prime\prime}\times1^{\prime\prime}})$ \\ 
 $m_i = C_i -2.5*\log(Flux^{ap}_i)-\delta m_i - k_i \times AM_i$; $i$ = $g'$, $b_{HIGH}$\\
 $k_i$ - atmospheric extinction coefficient  \\
 $\delta m_i$ - aperture correction 
 \end{tabular}
\label{t:mag}
\end{center}
\end{table}
 
\section{Discussion}
\label{sec4}

\begin{figure}
\setlength{\unitlength}{1mm}
\resizebox{12.cm}{!}{
\begin{picture}(120,113)(0,0)
\put (-2,0) {\includegraphics[width=9.1 cm,bb = 0 200 580 1000,clip=]{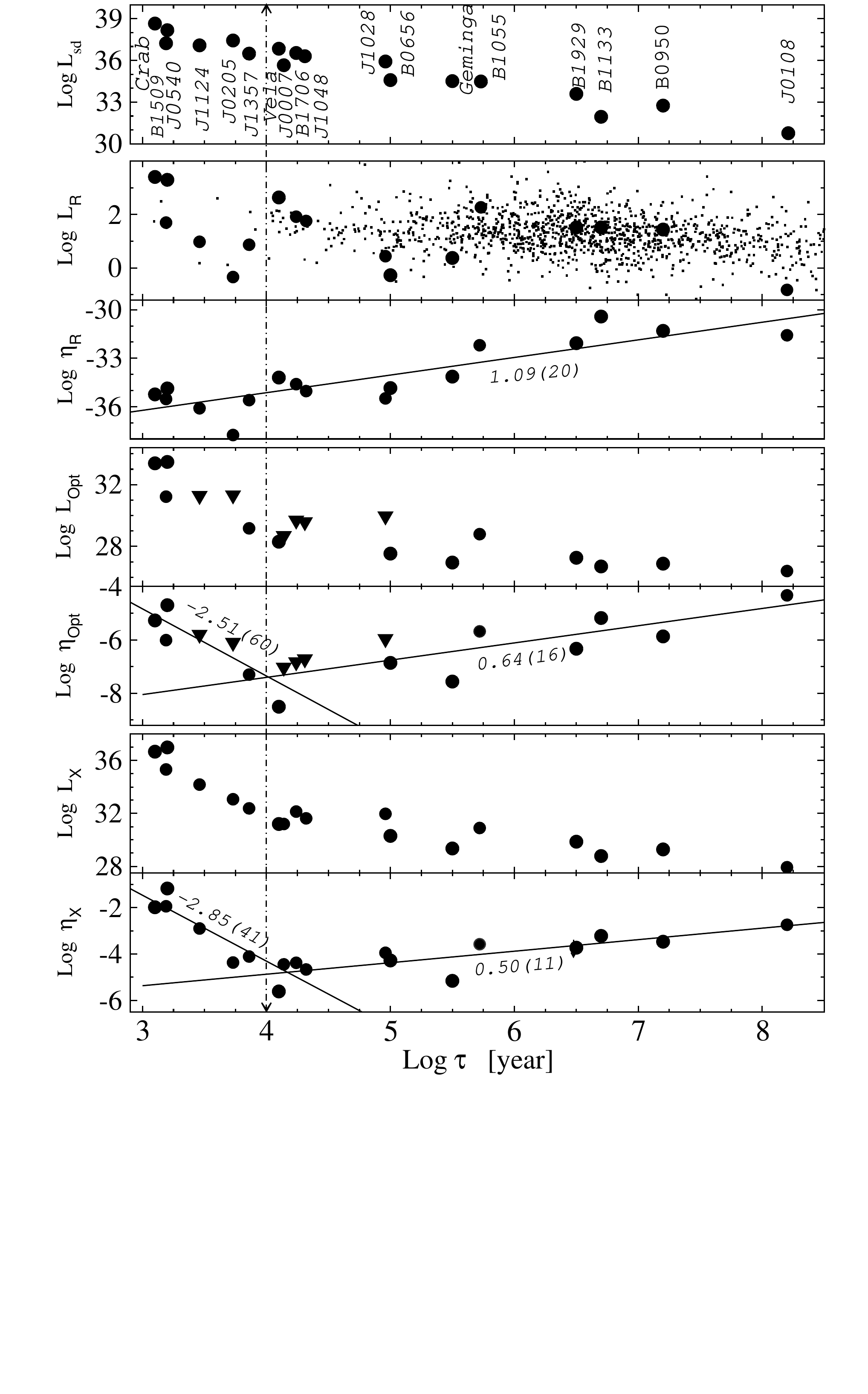}}
\end{picture}}
  \caption{From top to bottom: evolution of the pulsar spin-down, radio, optical and X-ray luminosity, and respective efficiencies, as a function of the dynamical age.  The full circles show the luminosity and radiation efficiency of  pulsars with optical counterparts, while the triangles  mark  upper limits. The points show the radio luminosity of the rest of radio pulsars taken from the ATNF pulsar catalogue (Manchester et al.\ 2005). The data  was taken from Zharikov et al.\ (2006) and updated using Danilenko et al.\ (2012), Mignani et al.\ (1999, 2009, 2010, 2011, 2013), Shibanov et 
al.\ (2008), Zharikov et al.\ (2008a, 2008b) and references there in. The vertical line marks the age when the change of the behaviour  of the radiation efficiency  occurs.  }
 \label{fig4}
\end{figure}

The measured fluxes of { the   candidate pulsar counterpart} in the three blue filters (2012 GTC $g'$, 2012 VLT $b_{\rm HIGH}$, and 2003 VLT $B_{\rm BESS}$) are consistent with each other (see Table~\ref{t:mag}).  Unfortunately, we could not obtain a more precise flux measurement because it is  close to  the detection limits of our new observations. However, we note that the { the   candidate pulsar counterpart}  does not  look redder in the $g'$ filter in comparison with the  $b_{\rm HIGH}$ and $B_{\rm BESS}$ ones,  although the low S/N ratio of the detection in each filter and their relatively close wavelengths do not allow us to draw firm conclusions on its spectrum.  This would exclude an extremely steep spectrum for { the   candidate pulsar counterpart},  which might have a flat spectrum like that of another similar-age pulsar PSR\, B0950+08 \citep{2004A&A...417.1017Z}.  

Our new detection  of the candidate counterpart would confirm that the relatively high values of the optical luminosity $\log L_B = 26.76(17)$ and efficiency $\log \eta_{Opt} = -5.2(2)$ of \psr\  are comparable to that of young and energetic Crab-like pulsars \citep{2008A&A...479..793Z}.  In Fig.~\ref{fig4} we show the   time evolution of the pulsar  radio ($L_R$),  optical ($L_{Opt}$) and  X-ray luminosity ($L_{X}$), and respective efficiencies $\eta_{R,Opt,X} = L_{R,Opt,X}/L_{sd}$,  as a function of the dynamical age\footnote{The spin-down luminosity $L_{sd} = 4\pi^2I\dot{P}/P^3$ and dynamical age $\tau = P/2\dot{P}$ are  combinations of the period and the period derivative and  we show $L_{sd}$ vs. $\tau$ for an illustration only.}($\tau$).   The data was taken from \citet{2006AdSpR..37.1979Z} and updated.  Eighteen pulsars with optical counterparts or with significantly deep upper limits on the optical luminosity are known.
  This sample of pulsars confirms the non-monotonic evolution of the pulsar  radiation efficiency in the optical and X-ray domains.
  While the spin-down luminosity obviously decreases with the pulsar dynamical age, the radio luminosity does not, which implies that the radio efficiency  must increase with the dynamical age. The observed dispersion in radio luminosity for pulsars of the same age likely results from a combination of several factors such as the magnetic inclination of their dipolar fields and viewing geometry of the radio emission beam.
  At variance with the radio efficiency,  there is a clear evidence of a change in the behaviour of the optical and X-ray efficiencies, which initially seem to decrease before starting to flatten or increase at larger ages. The turnover is located around $\tau\sim10^4$ years.  At the same time, the time evolution of $L_{opt}$, $L_{X}$ and $\eta_{opt}$, $\eta_{X}$ does not show any dependence on the pulsar surface magnetic field, which,   more likely, corresponds to an evolution of  the population of relativistic particles responsible for the emission.
We note that the  timescale $\tau\sim10^4$ years 
is  comparable to the transition between neutrino and photon cooling 
 stage \citep[and references therein]{1999PhyU...42..737Y, 2004AdSpR..33..523Y} in neutron stars.
 Therefore, we propose that the change of the cooling stage affects the distribution of relativistic particles in the pulsar magnetosphere, which is reflected in the dependence  of the optical/X-ray efficiency on the pulsar age.   The slopes of the time evolution of  $\eta_{opt}(\tau)$ and $\eta_{X}(\tau)$   after $10^4$~years are practically similar and  compatible with  that of $\eta_R(\tau)$ (Fig.~\ref{fig4}). The cooling theory predicts that after $10^5$ years the surface temperature of the neutron star drops below $10^5$ K and the emission is dominated by non-thermal radiation from the magnetosphere. Indeed, a non-thermal spectral component has been presumably detected in the broad-band optical emission from two nearby old pulsars: PSR\, B1929+10 \citep{2002ApJ...580L.147M} and PSR\, B0950+08 \citep{2004A&A...417.1017Z}. 

\section{Conclusion}

We observed the \psr\ field in the $B$ band in March 2012 with both the 10.4m GTC and the 8m VLT. We detected a faint  object, $g'=28.0(3)$ and $b_{\rm HIGH}=28.1(3)$, in both the GTC and the VLT images, whose position is coincident with the 2012 pulsar' radio coordinates computed after accounting for its proper motion  \citep{2002ApJ...571..906B}.   Although the object is at the detection limit in both the GTC and VLT images, its independent detection in the two data sets  suggests that it is real.   We detected no object at the position of the originally proposed optical counterpart of \psr\ \citep{2008A&A...479..793Z},  which was coincident with the 2003 pulsar's radio position. This would be naturally explained by the expected displacement of the proposed counterpart along the pulsar proper motion direction.   The flux of the object that we identified at the 2012 pulsar's radio position is also consistent with that of the proposed counterpart of \cite{2008A&A...479..793Z}, which was $B_{\rm BESS}=28.1(3)$. Therefore, we believe that we obtained a second-epoch detection of the pulsar candidate counterpart.  \psr\  would then be the eleventh  pulsar detected in the optical \citep{2011AdSpR..47.1281M}.   Confirming the identification of the optical counterpart of \psr\  would provide a possibility to study the broad-band energy distribution of another old pulsar, using the {\em Hubble Space Telescope} ({\em HST}) and/or large ground-based telescopes, and extend our ability in the understanding of non-thermal  magnetospheric emission of neutron stars.  In particular,  studying the pure non-thermal radiation from the magnetosphere of old pulsars can be a key to separate thermal and non-thermal radiation of younger pulsars.


\section*{Acknowledgments}
  The research leading to these results has received funding from the CONACYT 151858  project and the European Commission Seventh Framework Programme (FP7/2007-2013) under grant agreement n. 267251. We thank the anonymous referees for their constructive comments to our manuscript.

\bsp
\label{lastpage}
\end{document}